\documentclass{PoS}
\pdfoutput=1
\title{Prospects of New Physics Searches with the High Luminosity LHC}

\ShortTitle{Prospects of New Physics Searches with the HL-LHC}

\author{\speaker{Altan Cakir}\\
       {\rm On behalf of the CMS and ATLAS Collaborations}\\
       Deutsches Elektronen-Synchrotron (DESY)\\
       E-mail: \email{cakir@cern.ch}}

\abstract{After the observation of a Higgs boson near 125 GeV, the high energy physics community is investigating possible next steps for entering into a new era in particle physics. It is planned that the Large Hadron Collider will deliver an integrated luminosity of up to 3000/fb for the CMS and ATLAS experiments, requiring several upgrades for all detectors. The reach of various representative searches for supersymmetry and exotica physics with the upgraded detectors are discussed in this context, where a very high instantaneous luminosity will lead to a large number of pileup events in each bunch crossing. This note presents example benchmark studies for new physics prospects with the upgraded ATLAS and CMS detectors at a centre-of-mass energy of 14 TeV. Results are shown for an integrated luminosity of 300/fb and 3000/fb.}

\FullConference{Next Steps in the Energy Frontier - Hadron Colliders  \\
		 25-28 August, 2014\\
		 Fermilab, Batavia, USA}

\begin{document}

\section{Introduction}

The recent discovery of a Higgs boson marks a great triumph of experimental and theoretical physics. Nonetheless, the discovery has been only the first step in answering many unsolved questions in nature since many key elements are still left unexplained: We do not know how fundamental particles acquire their masses or why there exists dark matter in the universe. The Standard Model (SM) does not adequately explain these fundamental questions since it is inherently an incomplete theory. Therefore, over the last few decades, advanced developments in theoretical particle physics lead to important progress of fundamental physics phenomenology that can be probed at the Large Hadron Collider (LHC) frontier.  A wide range of searches for beyond the standard model (BSM) physics has been performed at the CMS and ATLAS experiments until now, unfortunately no evidence has been observed. 

Currently, the LHC is in the end of its first long shutdown (LS1) in order to prepare for running at $\sqrt{s}$ = $13$ TeV in $2015$, and a year later to the design energy of 14 TeV. The bunch spacing will most likely be expected to be 25 ns, the luminosity will reach the design value of 300/fb with 25 pile-up interactions by the end of 2021. A second long shutdown (LS2) in 2018 will take place to allow for an upgrade of the detectors for running at higher luminosities and an average pile-up of 50. The last phase of the planned LHC operation, referred to as the High Luminosity LHC (HL-LHC) \cite{hllhc}, will begin with the third long shutdown (LS3) in the period 2022-2023, where the machine and detectors will be upgraded to allow for proton-proton collision running at an average pile-up of 140. The ultimate goal will be an integrated luminosity of 3000/fb of data by the end of HL-LHC program. Solemnly, the increased pile-up environment in detectors degrades the ability to reconstruct quantities like the scalar sum of all jet momenta (HT) and the missing transverse momentum (MET), which are commonly used in most of BSM searches, and therefore different pile-up scenarios with various BSM analyses for the future HL-LHC should be investigated.

In this note, I present the results of sensitivity studies for searching for BSM models at the HL-LHC. The upgraded CMS and ATLAS detectors~\cite{jinst} are expected to collect 3000/fb of data with an average of 140 pile-up interactions per bunch crossing. In section 2, searches for Supersymmetry at the HL-LHC program are presented with various leptonic and hadronic final states~\cite{susy}. In section 3, analyses of vector boson scattering topologies with high-energetic forward jets are discussed~\cite{vecbos}. In section 4, searches for exotic models, i.e vector-like charge 2/3 quarks, search for $t \bar{t}$ and dilepton resonances and heavy stable charged particles,  are discussed ~\cite{exo} in the context of different signal topologies. All analyses are performed using parameterized fast detector simulations of the CMS and ATLAS \cite{atlcms} experiments in proton-proton collisions at center-of-mass energies of 14 TeV, corresponding to integrated luminosities of 300/fb and 3000/fb.

\section{Searches for Supersymmetry}

Supersymmetry (SUSY) is one of the best-motivated extensions of the SM for addressing most of its unsolved problems \cite{susyth}. SUSY is a proposed extension of space-time symmetry that relates two basic classes of elementary particles: bosons, which have an integer-value spin, and fermions, which have a half-integer spin. Each particle of the SM is associated with a particle from the other, called its super-partner, whose spin differs by a half-integer. SUSY provides elegant solutions to the unification of the gauge interactions and a radiative breaking of the electroweak symmetry. Under the conservation of R-parity, the lightest SUSY particle (LSP) is stable and is an excellent candidate for the dark matter in the universe. Therefore, the discovery (or exclusion) of weak-scale SUSY is one of the highest physics priorities for the current and future LHC programs, including the HL-LHC.  The section shows comparisons of the discovery and exclusion reach of R-parity conserving SUSY analyses with different hadronic and leptonic final states in the context of supersymmetric pair productions.

In the framework of generic R-parity conserving supersymmetric extensions of the SM, SUSY particles are produced in pairs and the LSP is stable. In a large fraction of the parameter space, the LSP is the lightest neutralino, where neutralinos ($\tilde{\chi}^{0}_{i}$, $ j = 1, 2, 3, 4$) and charginos ($\tilde{\chi}^{\pm}_{i}$ , $i = 1, 2$) are the mass eigenstates originating from the superposition of the SUSY partners of Higgs and electroweak gauge bosons (higgsinos and electroweak gauginos). The scalar partners of right-handed and left-handed fermions can mix to form two mass eigenstates, nearly degenerate in the case of first and second generation squarks and sleptons ($\tilde{q}$ and $\tilde{l}$). On the other hand, there is a possibility to split in the case of bottom and top squarks ($\tilde{b}$ and $\tilde{t}$) and tau sleptons ($\tilde{\tau}$ ). The lighter top squark mass eigenstate or the bottom squark can be significantly lighter than the other squarks and the gluinos (supersymmetric partners of the gluons). The following analyses are discussed in this context: Search for the direct production of gluinos and first and second generation squarks, search for the direct production of top and bottom squarks and search for the direct production of a neutralino and a chargino with the decay to a W or Z (Higgs) boson and the LSP.

\subsection{Strongly produced Supersymmetry: Searches for gluinos and the first and second generation of squarks}

\begin{figure}[b]
\centering
\includegraphics[width=5cm]{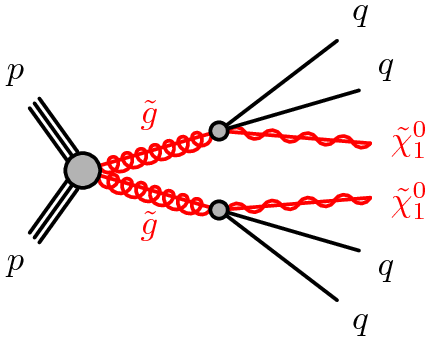}
\includegraphics[width=5cm]{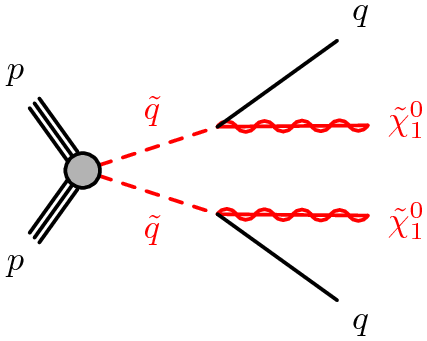}
\caption{The diagram for the gluino (left) and squark (right) pair production in simplified models. The gluino is assumed to decay into qq$\tilde{\chi}^{0}_{1}$ with $100\%$ branching ratio and the squark is assumed to decay into q$\tilde{\chi}^{0}_{1}$ with $100\%$ branching ratio. }
\label{fig:first}
\end{figure}

\begin{figure}[t]
\centering
\includegraphics[height=11cm]{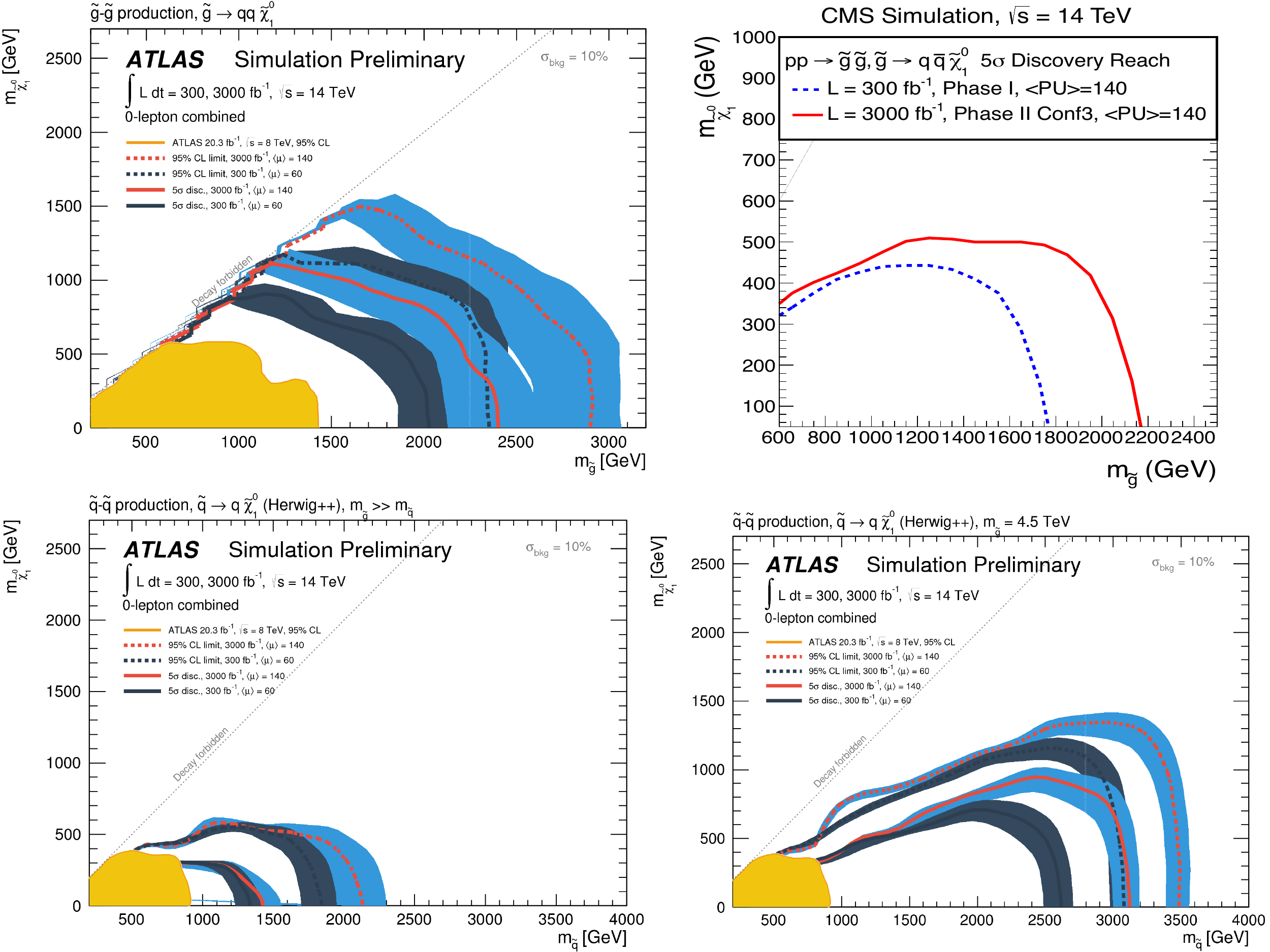}
\caption{Expected $95\%$ CL exclusion contours and 5$\sigma$ discovery contours for $L_{\mathrm int}$ = 300/fb and 3000/fb for gluino searches. In the CMS analysis, no exclusion limit has been shown (upper-right). For squark pair-production, the gluino mass is either (lower-left) decoupled or set to 4.5 TeV (lower-right). The bands reflect the 1$\sigma$ uncertainty on the production cross-section.  }
\label{fig:second}
\end{figure}

Strongly produced SUSY particles are expected to have the highest production cross-section of all SUSY processes, provided they are light enough to be produced at the LHC \cite{susyxs}. In this section, simplified models of gluino and squark pair production channels including up-to-date current exclusion limits from Run I are considered for the ATLAS and CMS experiments. The signal topologies in these studies are gluino and squark pair productions, in which each gluino or squark decays to one or two jets and the LSP as shown in Fig.~\ref{fig:first}. In both cases signal events are characterized by many jets, large MET and no leptons. 

The signal topologies are expected to produce a large amount of hadronic energy (HT), in association with different variables such as effective mass ($M_{eff}$), MET/$M_{eff}$ and MET/$\sqrt{HT}$. The signal regions for all topologies are similar to the selection in 8 TeV analyses for both collaborations. The SM background contribution to these channels generally arise from the following processes: Z($\nu$$\bar{\nu}$) + jets events, W(l$\nu$) + jets and $t \bar{t}$+ jets, where at least one W boson decays leptonically (l = e, $\mu$, or $\tau$). The W(l$\nu$) + jets events pass the search selection, when the e/$\mu$ escapes detection or a $\tau$ decays hadronically. QCD multijet events slightly contribute to the background because of jet energy mismeasurements or leptonic decays of heavy-flavor hadrons inside jets. The detail of applied selection cuts for these analyses can be found in the corresponding references.

Figure~\ref{fig:second} shows the expected exclusion and discovery contours for gluino and squark searches. The production cross-section for gluino searches is derived for the decoupling limit. Due to the lack of t-channel process contribution to the production cross section, the total cross-section for the limit calculation is affected much if squarks are as light as a few TeV.  As a result, the 5$\sigma$ discovery reaches for both CMS and ATLAS experiments are expected for gluino-pair production up to $m_{\tilde{g}}$ =~ 1800 to 2000GeV for 300/fb, and $m_{\tilde{g}}$ = ~2.3 TeV for the 3000/fb with increasing luminosity. Depending on the gluino mass, a $\tilde{\chi}^{0}_{1}$ with a mass up to 600-900 GeV (1100 GeV) can be discovered for the 300/fb (3000/fb) luminosity scenario. Notably, these mass ranges lie outside any current 8 TeV exclusion limits and differences for the ATLAS and CMS reaches come from different search bins between the analyses. The theoretical uncertainty on the ATLAS limit is mainly due to uncertainties in the parton-density functions and increases with gluino mass. The theoretical uncertainty reaches up to 400 GeV for gluinos with a mass of 2950 GeV. 
 
For squark-pair production two scenarios limit have been taken into account. In the first scenario, the squarks are considered as decoupled from the gluino. In the second scenario, the gluino mass is set 4.5 TeV, above the expected HL-LHC exclusion reach. Here, t-channel gluino exchange processes can significantly contribute and the production cross-section for heavy squarks is largely increased. If the gluino is decoupled, squarks can be excluded up to a mass of about 2000 GeV for light $\tilde{\chi}^{0}_{1}$. The limit is able to reach to 1850 GeV for an integrated luminosity of 300/fb. In this scenario, squarks can be discovered up to 1400 GeV with 3000/fb. If the gluino mass is close to the squarks, the discovery reach tends to be increased. For squark-pair production in the case where $m_{\tilde{g}}$ = 4.5 TeV, squarks can be discovered up to a mass of 2400 GeV for 300/fb and 3100 GeV for 3000/fb. 

\subsection{Third-generation SUSY: Direct stop and sbottom searches}

One example of a class of SUSY models that became important after the Higgs discovery has been labeled Natural SUSY or third-generation SUSY~\cite{natural}. Such models are called natural because a the superpartner of the top quark with a mass below one TeV is actually sufficient to keep the Higgs mass naturally light. Therefore, searches for the direct stop and sbottom pair productions have been carried out from both collaborations extensively last two years. 

\begin{figure}[b]
\centering
\includegraphics[height=4.8cm]{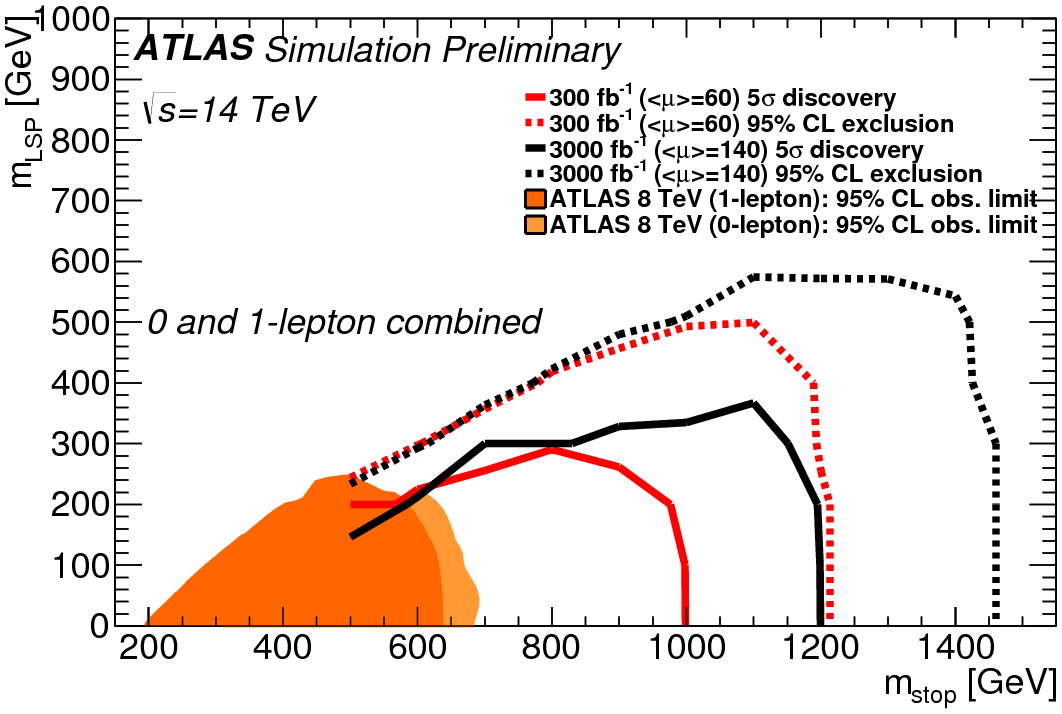}
\includegraphics[height=5cm]{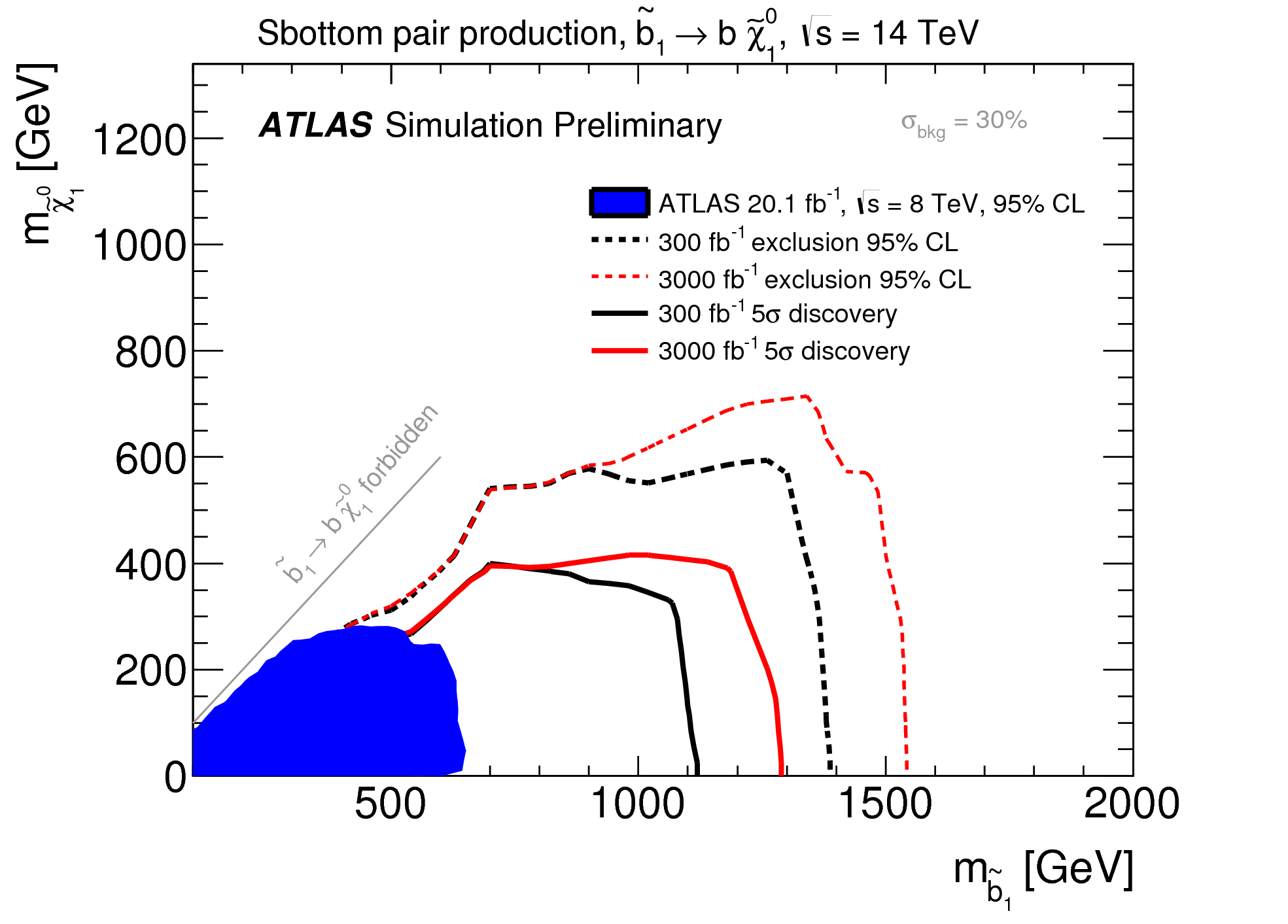}
\caption{Expected $95\%$ exclusion limits and discovery reaches for top and bottom squark pair productions with 300/fb and 3000/fb of integrated luminosity. }
\label{fig:third}
\end{figure}

The strategies pursued follow closely the 8 TeV analyses with re-optimization of the main variables. General signal topologies of such production channels, which are discussed in this subsection, include zero- and single- lepton final states. In addition, standard SUSY variables, such as HT, MET and MET/$\sqrt{HT}$ have been applied together with transverse and cotransverse related variables such as $M_T$ and $M_{CT}$~\cite{Tovey:2008ui}. $M_{CT}$, the main variable used to discriminate the sbottom signal, is calculated by an analytical combination of the heaviest and lightest SUSY particles masses ($\tilde{b}$ and $\tilde{\chi}^{0}_{1}$). 

Figure~\ref{fig:third} shows the discovery and exclusion potential for the combination of 1-lepton and 0-lepton analyses for stop searches at the HL-LHC. The signal and background events of the two analyses were added for the combination. For LSP masses below approximately 200 GeV a stop discovery at 5$\sigma$ would be possible with 3000/fb for stop masses up to approximately 1.2 TeV, assuming a $100$$\%$ branching ratio. The discovery reach is extended by approximately 200 GeV when increasing the integrated luminosity from 300/fb to 3000/fb. The 5$\sigma$ discovery reach and exclusion limits for sbottom searches are shown in Figure 13. Bottom squark masses up to 1.4 TeV can be excluded at $95$$\%$ CL with 300/fb of integrated luminosity, for a massles $\tilde{\chi}^{0}_{1}$. With 3000/fb at the HL-LHC, the exclusion reach improves by an additional 150GeV. Bottom squarks with masses of 1.1 TeV (1.3 TeV) can be discovered with 5$\sigma$ significance with 300/fb (3000/fb). 

\subsection{Electroweak production of SUSY particles: Direct Chargino and Neutralino production}

\begin{figure}[b]
\centering
\includegraphics[height=4cm]{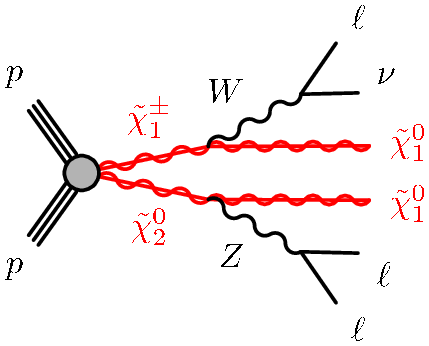}
\includegraphics[height=4cm]{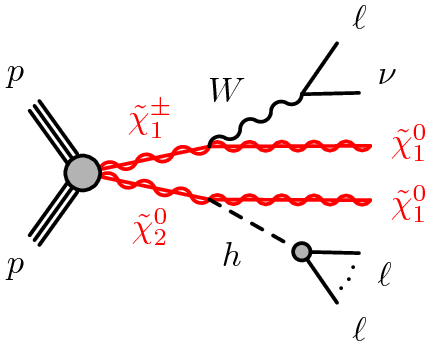}
\caption{The diagrams for the two $\tilde{\chi}^{\pm}_{1}$$\tilde{\chi}^{0}_{2}$ simplified models studied in this analysis. }
\label{fig:fourth}
\end{figure}

Based on naturalness arguments, the electroweaki sparticles are expected to have masses below $1$ TeV range and are potentially within the reach of the LHC. The cross section of the associated production of charginos and neutralinos ranges from 1000 to 1 fb for masses between 200 and 600 GeV and may contribute the SUSY production in scenarios with heavy squarks and gluinos. In this subsection, two simplified models of direct production of $\tilde{\chi}^{\pm}_{1}$ and $\tilde{\chi}^{0}_{2}$ are studied in this context, including WZ and Wh mediated channels. The relevant simplified models of these processes can be found in Figure~\ref{fig:fourth}. 

The final states can contain multi-leptons (3-leptons), where leptons refers to electrons and muons, including those from $\tau$-lepton decays but do not include hadronically-decaying $\tau$-leptons, and MET. In addition the single lepton plus two hadronically decaying taus ($\tau$) signature has been also used to study the Wh-mediated topology. The background for a signal with three leptons or a single lepton plus two $\tau$ leptons and large MET is dominated by the irreducible processes WZ, tribosons and $t \bar{t}$+ Z/W production. Events of the WZ channel are selected with exactly three leptons, and a same flavor opposite sign (SFOS) lepton pair is required to be present among the three leptons. Events with b-tagged jets are vetoed to suppress tt and tt + V (V= W, Z) backgrounds. Events are required to include at least one Z boson candidate, defined as a SFOS lepton pair with mass |$m_{SFOS}$ - $m_{Z}$| $<$ $10$GeV window. In the Wh channel, events are selected with exactly one lepton (electron or muon) and an opposite sign leptons, either electron and muon or $\tau$ pair. Details of the selections cuts for the WZ and Wh channels in multilepton searches can be found in the corresponding references. 

\begin{figure}[t]
\centering
\includegraphics[height=10cm]{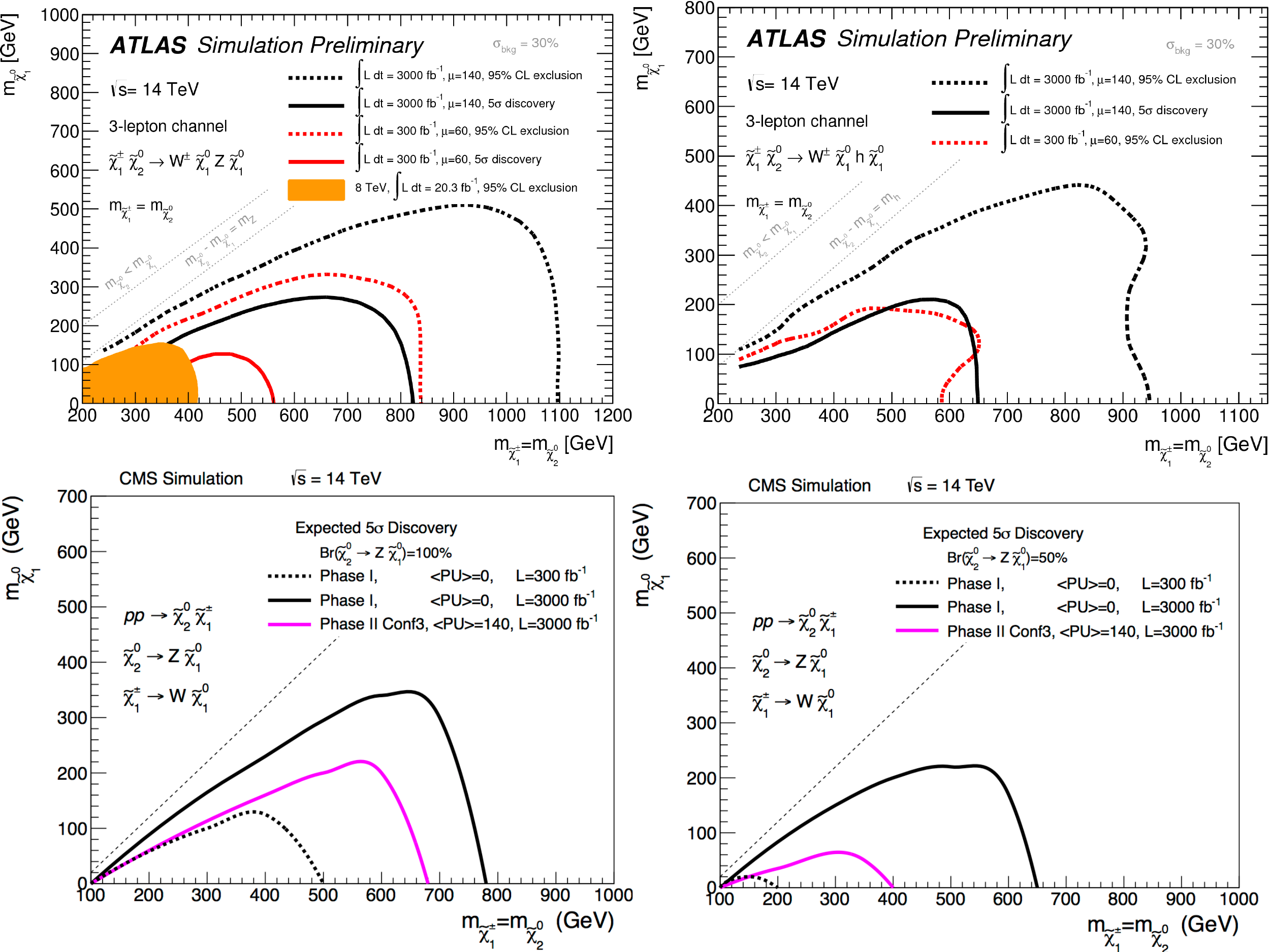}
\caption{The expected $95\%$ exclusion and discovery contours for the 300/fb and 3000/fb luminosity scenarios in the m($\tilde{\chi}^{0}_{1}$) vs m($\tilde{\chi}^{\pm}_{1}$ , $\tilde{\chi}^{0}_{2})$ plane for the WZ-mediated simplified model (upper-right and lower row) and Wh-mediated simplified model limit (upper-right), which higgs decays to light leptons (e, mu) has been considered.}
\label{fig:fifth}
\end{figure}

The $95$$\%$ exclusion and 5$\sigma$ discovery contours for the WZ-mediated simplified models can be seen in Figure~\ref{fig:fifth}. In the case of the WZ-mediated topology, the exclusion contour reaches 840 GeV in $\tilde{\chi}^{\pm}_{1}$ and $\tilde{\chi}^{0}_{1}$ mass for 300/fb scenario, while the contour reaches around 1.1TeV in $\tilde{\chi}^{\pm}_{1}$ and $\tilde{\chi}^{0}_{1}$ mass for the HL-LHC scenario.The discovery contour reaches 560 GeV and 820 GeV in $\tilde{\chi}^{\pm}_{1}$ and $\tilde{\chi}^{0}_{1}$ for the luminosity scenarios with 300/fb and 3000/fb, respectively.

\section{Vector Boson Scattering and Triboson Production}

Signal topologies of the vector boson scattering and triboson production channels at the HL-LHC have been considered. The new interactions that can produce diboson pairs are determined by the gauge structure of the SM. An important feature of these production channels is the presence of two high-pT jets in the forward regions and different hemispheres, similar to those found in Higgs production via vector boson fusions. The absence of color exchange in the hard scattering process may produce a rapidity gap in the central part of the detector. However, this gap topology may lead to difficulties due to the high level of pileup at the HL-LHC. 

\begin{figure}[t]
\centering
\includegraphics[height=2cm]{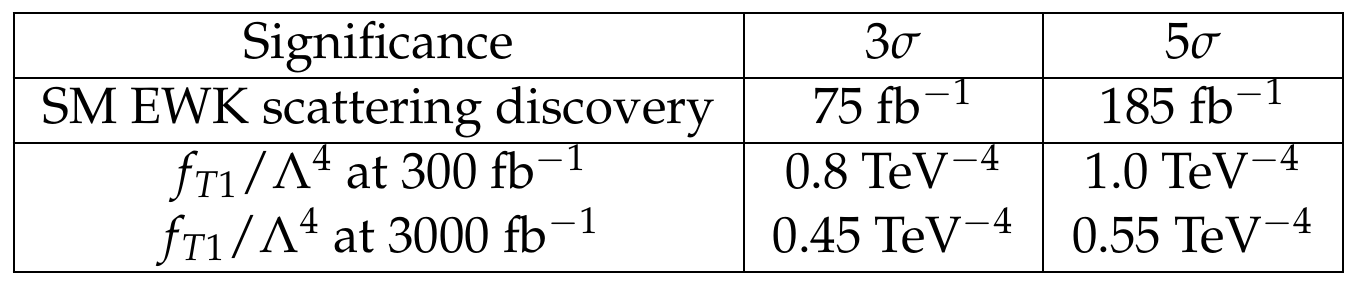}
\caption{Sensitivities for SM EWK scattering discovery and aQGC from the CMS experiment. The integrated luminosities for SM EWK discovery at 3$\sigma$and 5$\sigma$ are reported while aQGC prospects for discovery are given in terms of the operator coupling constant $f_{T1}/\Lambda^{4}$}
\label{fig:tb1}
\end{figure}

\begin{figure}[t]
\centering
\includegraphics[height=3.2cm]{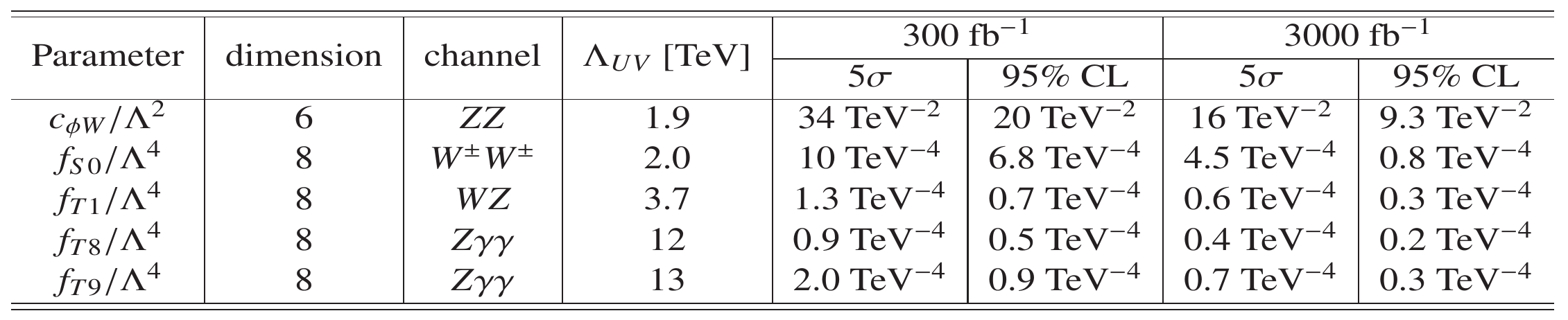}
\caption{$5\sigma$-significance discovery values and $95\%$ CL limits for coefficients of higher-dimension operators from the ATLAS experiment. $\Lambda_{UV}$ is the unitarity violation bound corresponding to the LHC and HL-LHC programs}
\label{fig:tb2}
\end{figure}

The nature of the electroweak sector predicts the self-interaction of electroweak gauge bosons in the form of triple and quartic gauge boson couplings, called as TGC and QGC respectively. Although the study of these couplings offers an important check of the SM precision tests, QGCs are additionally connected to the electroweak symmetry breaking sector, together with the Higgs boson, to ensure unitarity at high energies for scattering processes. Thus, QGCs offer a new tool on the electroweak symmetry-breaking mechanism and represent a sector for observing new physics due to anomalous couplings. Therefore, new physics can be parameterized in terms of higher-dimension operators in an effective field theory since the SM Lagrangian contains dimension-$4$ operators. Multi-boson production in the new physics context can be parameterized by certain dimension-$6$ and dimension-$8$ operators containing the Higgs and/or gauge boson fields. 

The main interest in these analyses is on the coefficient $c_{\phi W}$/$\Lambda^{2}$~\cite{cphi}, which is associated with a dimension-6 operator, as well as $f_{S0}$/$\Lambda^{4}$, $f_{T1}$/$\Lambda^{4}$, $f_{T8}$/$\Lambda^{4}$ and $f_{T9}$/$\Lambda^{4}$~\cite{TLam}, which are coefficients associated with dimension-8 operators. Here $\Lambda$ is a mass-dimensioned parameter associated with the new physics. The following TGC and QGC analyses are considered for the different final states. The fully-leptonic ZZjj  channel has a small cross section but provides a clean, fully reconstructible ZZ final state. On the other hand, the fully leptonic WZjj (as well as WWjj) channel has a larger cross section than ZZjj and can still be reconstructed by solving for the neutrino momentum using the W boson mass constraint. A forward jet-jet mass requirement of 1 TeV for all topologies reduces the contribution from jets accompanying non-VBS diboson production for all production channels. In the triboson channel, the Z$\gamma \gamma$ mass spectrum at high mass is sensitive to BSM contributions. Therefore, the lepton-photon channel allows full reconstruction of the final state and calculate the Z$\gamma \gamma$ invariant mass. In addition, the $10$ GeV invariant mass window requirement around the Z  boson mass peak can suppress the $\gamma^{*}$ contribution to the dilepton. The large angle requirement between photon and lepton and the high transverse momentum requirement of the photon reduces the FSR contribution. This leads to the phase space which is uniquely sensitive to the QGC. In tables~\ref{fig:tb1} and~\ref{fig:tb2}, CMS and ATLAS results for different dimension-6 and dimension-8 operators in the context of the relevant paramaters and production channels are presented. 

\section{Search for Exotic Models}

The HL-LHC substantially increases the potential for the discovery of exotic new particles. The range of possible phenomena is quite large. In this section I discuss some exotic models of BSM physics and the expected gain in sensitivity from the order of magnitude increase in integrated luminosity provided by the HL-LHC. 

\subsection{Vector-Like Charge 2/3 Quark Search}
\begin{figure}[b]
\centering
\includegraphics[height=5cm]{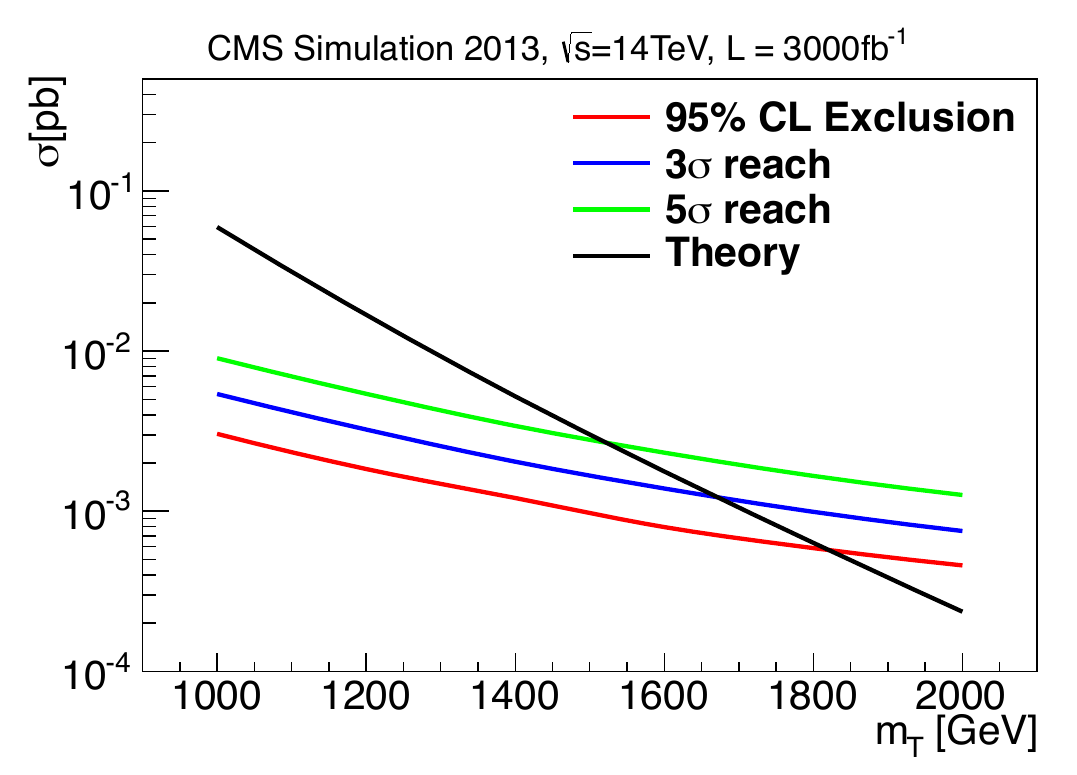}
\caption{Expected sensitivity for a T quark pair production signal in the combined multilepton and single-lepton+jets sample. The $95\%$ C.L. limits, $5\sigma$ discovery reach and the $3\sigma$ discovery reach are shown as a function of the T quark mass.}
\label{fig:sixth}
\end{figure}

Electroweak couplings of vector-like quarks differ from SM quarks. In the SM, quarks have a V-A coupling to the W boson leading to a different coupling of the left-right states. However, vector-like quarks phenomenologically have only vector-couplings to the W boson. Therefore, one can write a mass term for them that does not violate gauge invariance without the need for a Yukawa coupling to the Higgs boson. Vector-like quarks are extensively discussed in phenomenology~\cite{vector}. In this model, massive quarks, T quarks, have been predicted in the context of phenomenological approaches. Events from the decay of massive T quarks are characterized by two to four vector bosons and at least two b quarks at the LHC. Depending on the decay of the vector bosons, the final states can have varying numbers of isolated leptons, namely single lepton and multilepton channels. 

The $95$$\%$ C.L. exclusion reach is expected to be around $1.85$ TeV in Fig.~\ref{fig:sixth} at the HL-LHC, a factor of $2.5$ larger compared to the current LHC limit. This result would have a significant impact on the implications of a light Higgs at $126$ GeV on composite Higgs models in future, since light top partners with masses around a few TeV are essential to understand for a moderate level of tuning to the Higgs boson mass.

\subsection{Search for $t \bar{t}$ and Di-lepton Resonances}

\begin{figure}[b]
\centering
\includegraphics[height=11cm]{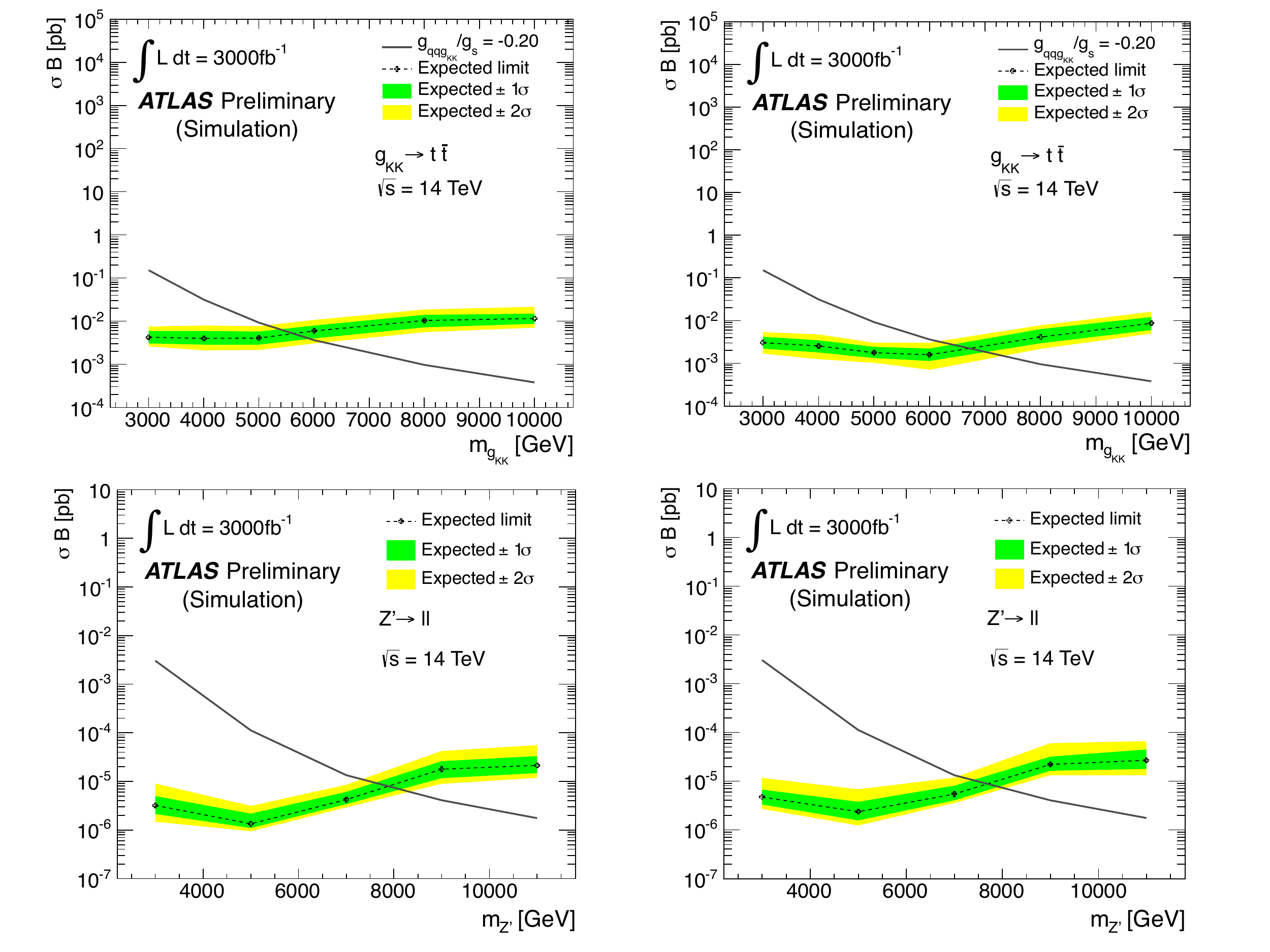}
\caption{The reconstructed mass limit for $g_{\mathrm KK}$ $\rightarrow$ $t \bar{t}$ in the dilepton (upper-left) and lepton+jets (upper-right) channels with 3000/fb. The reconstructed dielectron (lower-left) and dimuon (lower-right) resonance limits for the $Z^{\prime}_{SSM}$ search with 3000/fb. }
\label{fig:seventh}
\end{figure}

The BSM models with $t \bar{t}$ resonances provide benchmarks for cascade decays containing leptons, jets (including b-quark jets) and MET due to the nature of signal topology. In addition, it also provides the opportunity to study highly boosted topologies at higher energies. The sensitivity to the Kaluza-Klein gluon ($g_{\mathrm KK}$) via the process pp $\rightarrow$ gKK $\rightarrow$ $t \bar{t}$ and a heavy $Z^{\prime}$ decaying to $t \bar{t}$ in both the dilepton and the lepton+jets decay modes at the HL-LHC are considered in this context. 

For the sensitivity to a $Z^{\prime}$ boson or dilepton resonances, the dielectron and dimuon channels are taken into account separately. In these resonances, momentum resolutions scale differently with $p_T$ and the detector acceptances are different for electrons and muons. The  background is dominated by the SM Drell-Yan production, while $t \bar{t}$ and diboson backgrounds are substantially smaller. 

The statistical analysis is performed by a likelihood fit of templates of these distributions, using background plus varying amounts of signal, to the simulated data. The resulting limits as a function of the $g_{\mathrm KK}$ pole mass for the dilepton and lepton+jets channel are shown in Fig.~\ref{fig:seventh} (upper row), respectively.  The resulting limits as a function of the  $Z^{\prime}$ pole mass for the dielectron and dimuon channels are shown in Fig.~\ref{fig:seventh} (lower row), respectively.

\subsection{Search for Heavy Stable Charged Particles}

\begin{figure}[b]
\centering
\includegraphics[height=11cm]{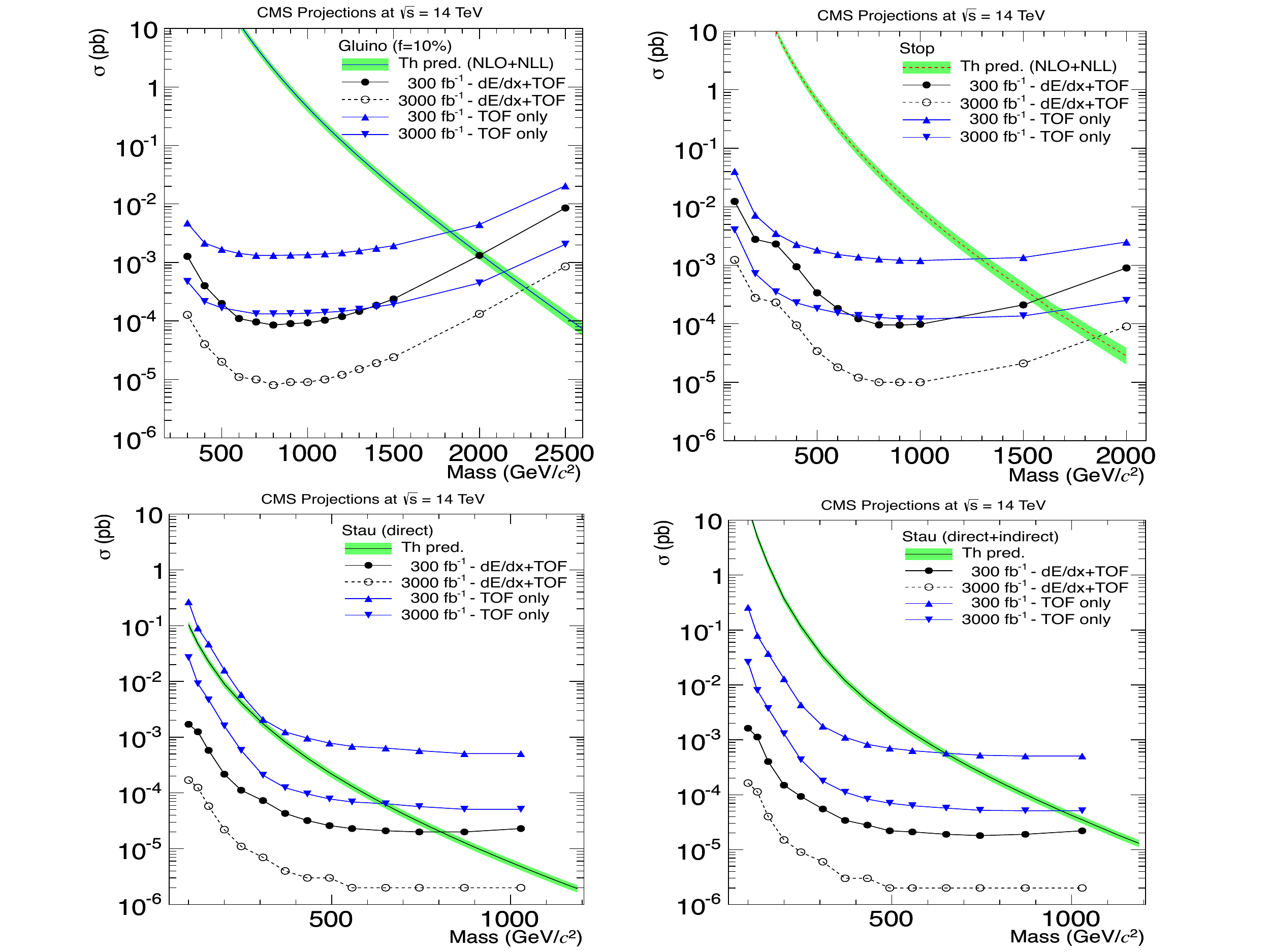}
\caption{Minimum cross sections for an expected signal significance of 5$\sigma$ standard deviations. The signal models considered are the pair production of gluinos (upper-left) and of stops (upper-right). The signal models considered are the direct pair production of staus (lower-left) and direct+indirect production of $\tilde{\tau}$`s (lower-right) in the context of GMSB.}
\label{fig:acht}
\end{figure}

These searches in the CMS Collaboration present the most stringent limits to date on long-lived gluinos, scalar top quarks, and scalar $\tau$ leptons. The signatures include long time-of-flight to the outer muon system and anomalously large energy deposition in the inner tracker of the CMS detector. The existing results are presented for each channel separately and in combination. Unlike many conventional searches at the LHC, where backgrounds arise from irreducible physical processes, the background to these signal topologies comes primarily from detector systems. It is therefore assumed that the backgrounds scale linearly with integrated luminosity, resulting in a constant signal over background ratio. By scaling the signal yields linearly with integrated luminosity from the 8 TeV result, a conservative assumption about the signal acceptance is introduced, since 14 TeV kinematics are expected to yield increased acceptance.

Several changes are accounted for the HL-LHC and detector operating conditions anticipated conditions after the CMS detector upgrade. First, since the LHC has operated at 50 ns bunch spacing during the Run I, the 8 TeV search was able to utilize a wide muon trigger time window, accepting candidates that arrive one LHC bunch-crossing after the collision. The LHC is expected to run with 25 ns bunch-spacing from 2015 onwards, resulting in a reduced trigger time window, so the signal efficiency used in these projections has been adjusted, based on fully simulated 8 TeV Monte-Carlo events. Secondly, the current dE/dx measurement relies on analog readout of the CMS Tracker, which will almost certainly not be possible after the CMS Tracker is upgraded during LS3. To account for this, the sensitivity with 3000/fbis presented for two scenarios: the combination of long time-of-flight and highly ionizing signatures, corresponding to an assumption that the dE/dx performance remains unchanged, and the sensitivity using the long time-of-flight signature alone, corresponding to an assumption that dE/dx measurements cannot be performed with the upgraded CMS Tracker.

The results in Fig.~\ref{fig:acht} shown that the additional integrated luminosity will lead to sensitivity for long-lived particles produced with cross sections at least one order of magnitude lower than what has been excluded by the current analysis. The models considered in this search are simple benchmarks and the search for long-lived particles even in the already excluded mass range is considered. While the cross sections of the relevant BSM models are model dependent, the analysis itself is signature-based and mostly decoupled from any given theoretical model.  

\section{Conclusion}

The discovery of the Higgs boson is the most important result from the Run 1 for understanding the fundamental laws of physics. There are many reasons to believe that more discoveries are imminent at the upcoming runs and the HL-LHC program. The rich BSM physics program, shortly summarized in this note, provides strong justification for the upgrade programs of the CMS and ATLAS detectors to exploit these magnificent and unique opportunities.

\end{document}